\documentclass[conference]{IEEEtran}

\usepackage{multirow}

\ifCLASSINFOpdf
  \usepackage[pdftex]{graphicx}
\else
\fi

\begin{document}

\title{A Study of Energy and Locality Effects\\using Space-filling Curves}
\author{\IEEEauthorblockN{Nico Reissmann\\ and Magnus Jahre}
\IEEEauthorblockA{
Dept. of Computer and Information Science (IDI)\\
NTNU, Norway\\
\{reissman, jahre\}@idi.ntnu.no}
\and
\IEEEauthorblockN{Jan Christian Meyer}
\IEEEauthorblockA{
HPC Section, IT Dept.\\
NTNU, Norway\\
Jan.Christian.Meyer@ntnu.no}
}

\maketitle

\begin{abstract}
The cost of energy is becoming an increasingly important driver for the operating cost of HPC
systems, adding yet another facet to the challenge of producing efficient code.
In this paper, we investigate the energy implications of trading computation for locality using 
Hilbert and Morton space-filling curves with dense matrix-matrix multiplication. The advantage of
these curves is that they exhibit an inherent tiling effect without requiring specific architecture
tuning. By accessing the matrices in the order determined by the space-filling curves, we can trade
computation for locality. The index computation overhead of the Morton curve is found to be balanced
against its locality and energy efficiency, while the overhead of the Hilbert curve outweighs its
improvements on our test system.
\end{abstract}

\IEEEpeerreviewmaketitle

\section{Introduction}
Cooling systems and technology scaling challenges make power consumption an increasingly important
cost factor in HPC systems \cite{Feng07}, with estimates of approximately 50\% of total
annualized system cost reported already in 2009 \cite{Koomey09}. Techniques such as clock
gating and dynamic voltage and frequency scaling (DVFS) have mitigated this development
architecturally, but software developments are not expected to contribute similarly
\cite{Dawson-Haggerty09, Seng03}. This suggests that with present systems the optimization for
program performance is equivalent with achieving energy efficiency \cite{Seng03, Yuki13, Pallister13}.

Processor and memory subsystems account for the two biggest components in a total system
power budget \cite{Bircher07}, and program memory access patterns correlate both with power and energy.
Specifically, higher spatial and temporal locality increase power consumption, but also improve energy
efficiency \cite{Song09} by reducing the energy lost waiting for memory operations and improving the
execution time.

For matrix manipulation, tiled algorithms improve utilization of the memory hierarchy
by increased reuse of submatrices fetched into cache memory, but their effectiveness is highly
dependent on cache parameters and problem size \cite{Lam91}.

\begin{figure}
	\centering
		\includegraphics[scale=0.5]{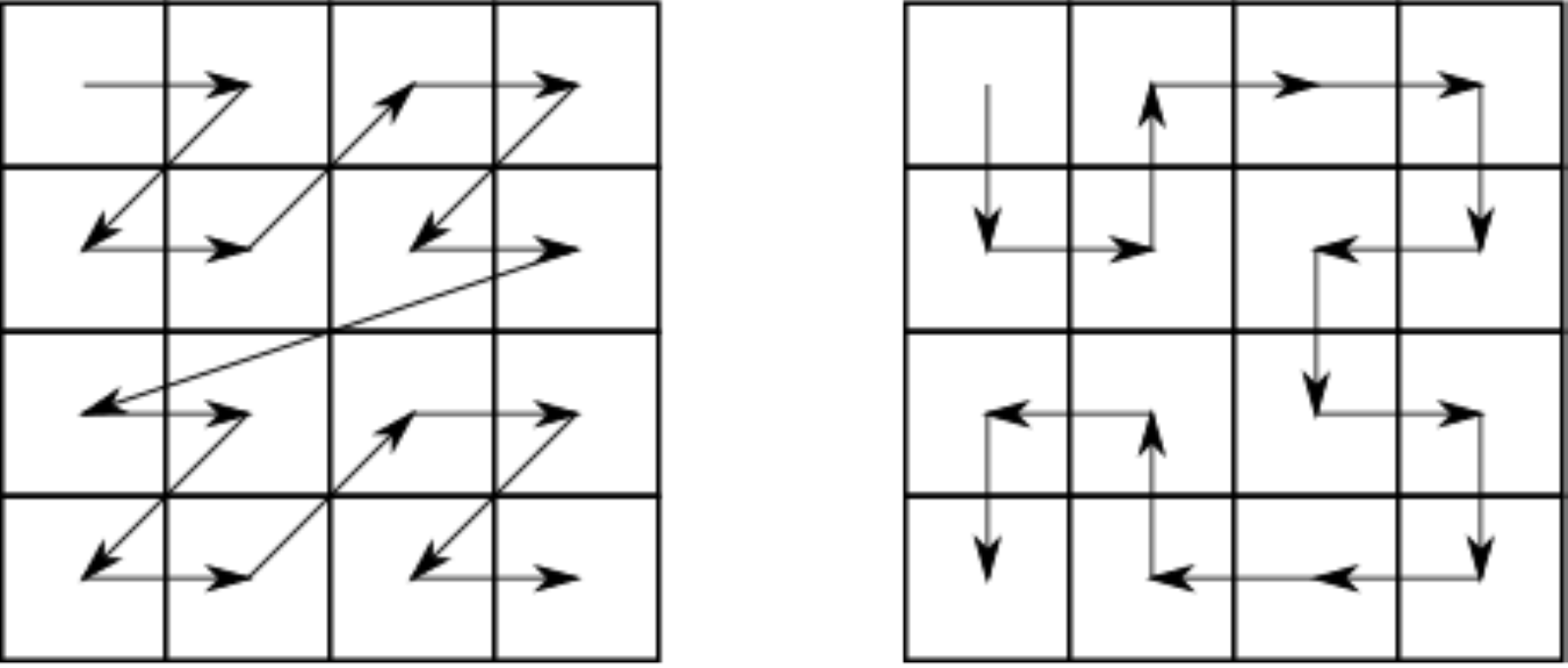}
	\caption{Traversal of $4\times 4$ matrices in Morton and Hilbert orders}
	\label{fig:sfcurves}
\end{figure}

Another way to improve spatial locality is to alter the ordering of matrix elements in memory
according to a space-filling curve that implies a multi-level tiling pattern. This type of
technique does not achieve the performance levels attainable by explicitly tiled, automatically
tuned approaches like ATLAS \cite{Whaley}, but is favorable in contexts where cache-oblivious
behavior is desired. Examples include software deployed as identical binaries on multiple
target architectures, or when the additional computational effort of automatically tuning
an installation is impractical. Two well known curves that exhibit an inherently
tiled access pattern are the Morton and Hilbert curves, shown for $4\times4$ matrices in Figure
\ref{fig:sfcurves}.

The advantage of reducing the dependency on architecture-specific parameters comes at
the expense of increased computational cost, because calculating the memory address of
an element from its matrix coordinates in these orderings requires computations which
take logarithmic and linear time with respect to address length, as opposed to the
constant cost of conventional row- or column-major orders. Motivated by the optimization
guideline that memory operations can be orders of magnitude more expensive than computation on
recent and future architectures \cite{Asanovic}, the recent addition of
energy counters in mainstream architectures \cite{David10} enables
us to empirically investigate the energy impact of this trade-off between
general and architecturally specialized methods in an energy-efficiency context.

In this paper, we apply the Morton and Hilbert orderings to general matrix-matrix
multiplication, and compare their utility to naive row-major indexing, in order to
quantify their impact on computation time and energy consumption.
We find that both curves improve spatial locality, but that the Hilbert ordering
replaces the avoided memory accesses with an amount of computation that negates
its utility on our test platform.

The rest of the paper is organized as follows:
Section \ref{sfcurves} discusses the procedure of constructing Morton and Hilbert curves,
and relates their definitions to the complexity of computing memory addresses
from coordinate pairs in both orders, Section \ref{methodology} describes our test
platform, instrumentation methods and experimental setup, Section \ref{results} 
summarizes and discusses the resulting measurements, Section \ref{relwork} discusses
related work, and Section \ref{conclusion} concludes with a summary of our
findings, and suggestions for future investigation.

\section{Space-filling Curves}
\label{sfcurves}

\begin{figure}
\centering
\includegraphics[scale=0.5]{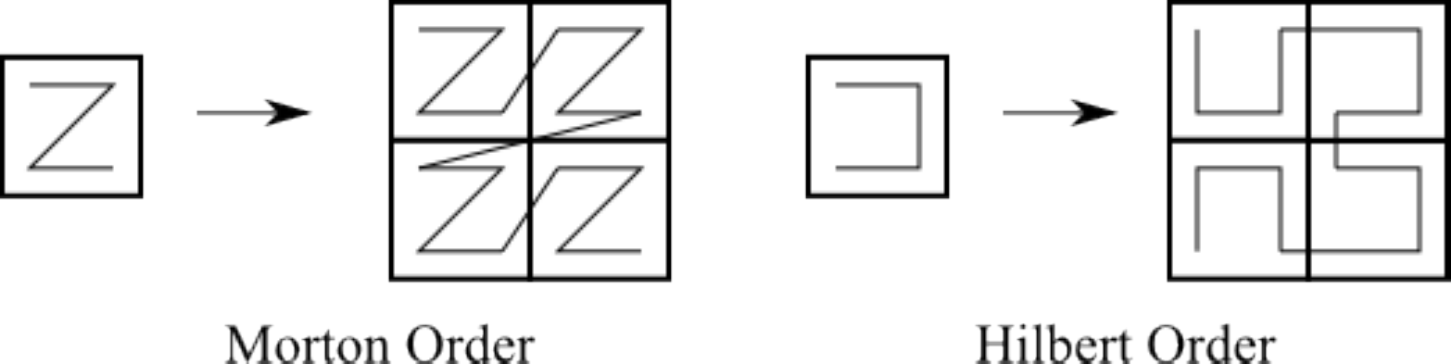}
\caption{Inductive steps of Morton and Hilbert curve constructions}
\label{fig:induction}
\end{figure}

Informally, the Morton and Hilbert curves are inductively constructed from basic
$2\times 2$ squares by replication and rotation, as illustrated in
Figure \ref{fig:induction}. In order to use these curves as encodings of matrices in a linear memory
array, procedures for mapping the two-dimensional indices onto one-dimensional memory arrays are
required. The remainder of this section outlines these procedures and the computational complexities
of their algorithms. 

\subsection{Serialization of Quadrant Indices}

The inductive constructions of the Morton and Hilbert curves share the feature that
both are defined by a traversal order of the 4 quadrants in a square.
This suggests that the indexing scheme to serialize matrix elements in a linear
memory array should similarly partition the array into quarters, making
the memory address of an element equivalent to a sequence of successively refined
quadrant selections. Each 4-way selection is thus encoded as a pair of bits.

\begin{figure}
\centering
\includegraphics[scale=0.5]{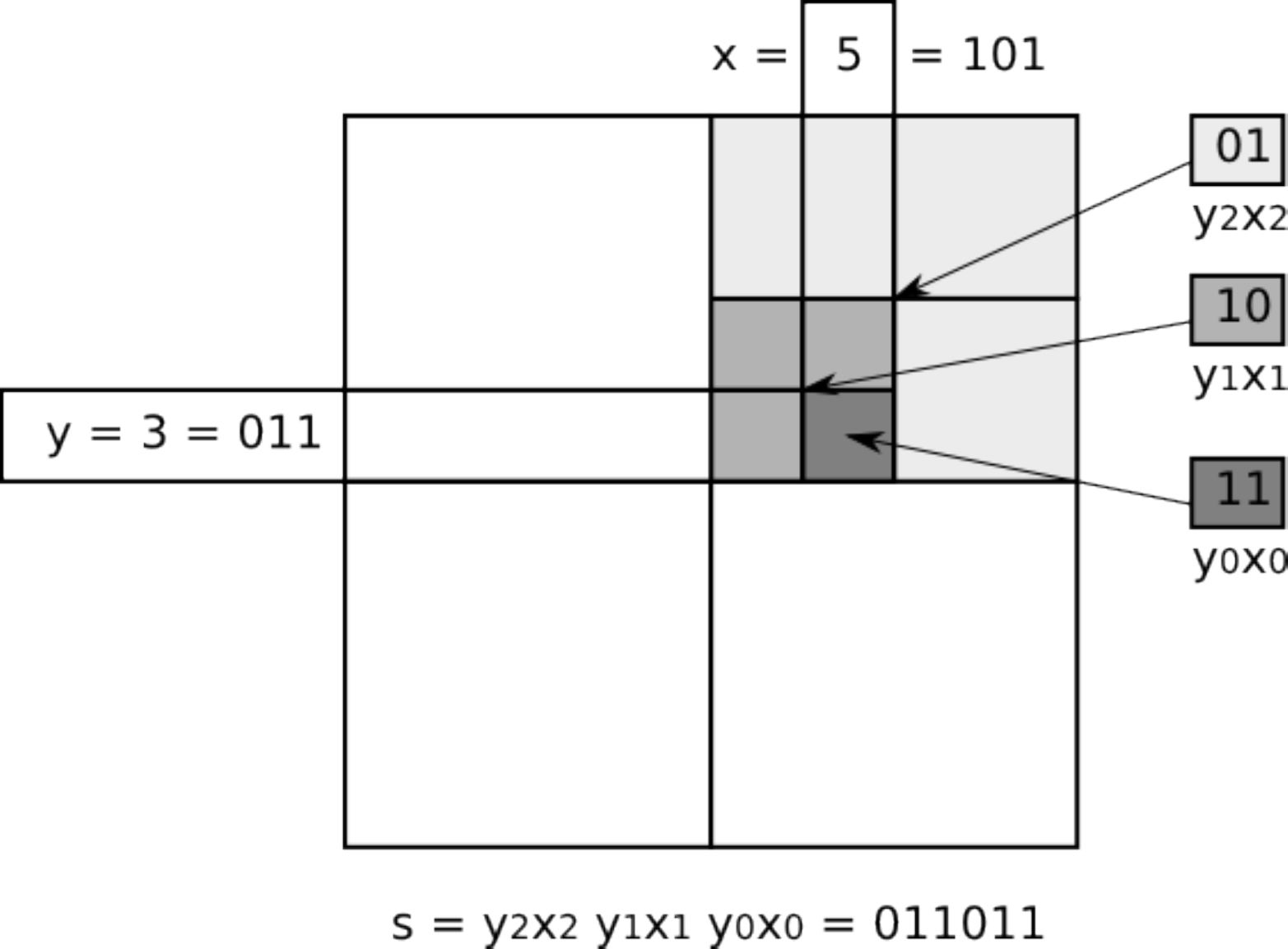}
\caption{Serialization of indices y=3 and x=5 by bitwise interleaving}
\label{fig:quadrants}
\end{figure}

The serialized index of a coordinate pair $(y,x)$ with y as the
major coordinate can be found by a bitwise interleaving of the coordinate pair,
combining the most significant bit of both coordinates to select which $\frac{1}{4}$
partition of the array to address, the next pair to select which $\frac{1}{16}$ part
within it, and so on. Figure \ref{fig:quadrants} illustrates the mapping for the
coordinate pair $(y=3,x=5)$, with y representing the major coordinate, and varying
vertically in the figure.

Interleaving two arbitrary-length bit patterns asymptotically requires computation
time linearly proportional to their length. Observing that these bit patterns
represent memory addresses, we restrict our considerations to lengths that fit
within address registers of a target machine, providing constant-time
operations to combine pairs of binary strings. Moreover, we note that as the
serialized bit pattern $s$ is as long as the sum of its constituent
coordinates, this restricts coordinates to half the size of machine registers,
admitting pairs of 32-bit coordinates on a 64-bit architecture. As 32-bit
coordinates provide ample space for matrix sizes that outgrow the last-level
caches of modern processors, we will assume that register level operations can
be used to interleave them, and sacrifice some generality to the advantage
of assuming that bit patterns can be dilated and combined in pseudo-logarithmic
time with respect to their length. Dilation of integers has been thoroughly
studied in the literature \cite{Stocco95}. We use an algorithm due to
Raman and Wise \cite{Raman08} which our implementations adopt as a constant
sequence of 5 shifting and 5 masking operations, involving 5 constant values
and 1 register.

\subsection{Morton and Hilbert Orders}

\begin{table}[!t]
\caption{Morton (MO) and Hilbert (HO) Traversal Orders}
\label{tab:traversals}
\centering
\begin{tabular}{|c|c|c|}
\hline
{\bf MO} & {\bf 0} & {\bf 1}\\
\hline
{\bf 0} & 0 & 1\\
\hline
{\bf 1} & 2 & 3 \\
\hline
\end{tabular}
\hspace{24pt}
\begin{tabular}{|c|c|c|}
\hline
{\bf HO} & {\bf 0} & {\bf 1}\\
\hline
{\bf 0} & 0 & 1\\
\hline
{\bf 1} & 3 & 2\\
\hline
\end{tabular}
\end{table}

Although the serializations of Morton and Hilbert indices work according to the
same recursive decomposition of a 2-dimensional area, they are differentiated by
applying different traversal orders of the quadrants at each level.
Table \ref{tab:traversals} lists their respective traversal sequences.
Note that the Morton order essentially applies the same straightforward ordering 
as row-major order, but does so recursively, which gives a tiling effect
that improves spatial locality. As this pattern essentially amounts to following
the same order as conventional binary counting, interleaving the major and minor
coordinates already completes the index translation
for this scheme. The resulting index calculation cost is higher than row-major
ordering, but remains constant for indices within register size, at the cost of
introducing minor discontinuities between quadrants $(1,2)$ and $(3,4)$, and
a larger gap between quadrants $(2,3)$, as was already seen in
Fig. \ref{fig:sfcurves}.

The Hilbert order eliminates these gaps by following a traversal order that
also steps between neighboring elements across quadrant boundaries. This
results in further improvement of spatial locality, but it comes at an
additional computational cost, as interleaved coordinates must be further
rearranged by additional computation, because the traversal orders for quadrants
rotate as a function of their position and depth in the recursive decomposition.

Lam and Shapiro \cite{Lam91} give an iterative description of this function which
scans coordinate bit pairs, and produces the rotation as a series of swap and bitwise
complement operations applied to the bits trailing the examined pair. This
adds a linear term to the overall complexity of the index serialization.

\section{Methodology}
\label{methodology}

\subsection{Execution Platform}
We conduct our experiments on an Intel Sandy Bridge architecture with the specifications shown
in Table \ref{tab:specs}. The processors feature \textit{
Running Average Power Limit} (RAPL) Model Specific Registers \cite{David10}, which gather the energy
consumed by the CPU package and memory module. By default, these
performance counters provide estimates of consumed energy in multiples of $15.3\mu J$. We obtain
estimates using the PAPI 5.3.0 library\footnote{http://icl.cs.utk.edu/papi/}, which provides a
high-level interface for reading performance counters.

\begin{table}[!t]
\caption{Test System Specifications}
\label{tab:specs}
\centering
	\begin{tabular}{|c|c|}\hline
	Processor & 2 E5-2670 2.6 GHz Xeon processors\\
						& 8 cores and 16 threads per processor\\
						& (32 threads with hyperthreading)\\
	Cache			& 20 Mb shared L3-cache\\
	Memory 		& 8x8Gb at 1600MHz\\\hline
	\end{tabular}
\end{table}

\subsection{Software Implementations}
The experimental programs are written in C, parallelized using OpenMP, and compiled
using gcc 4.7.2, with {\tt -O3 -mavx -fopenmp} optimizations enabled.
In dual socket experiments, all threads are evenly distributed between the sockets.
Power estimates are derived from periodically measuring RAPL Model Specific Registers at a rate
of 10Hz and checked against the power measurements of a Yokogawa WT210 power measurement device
operating at the same sampling rate. Energy estimates are obtained from the power logs through
numerical integration, by applying the trapezoidal rule. The intervals of the time integration were
obtained from the timestamps of the power estimates.

\subsection{Experimental Procedure}
Table \ref{tab:experiments} gives an overview of our experiments. We implemented
the naive matrix multiplication for three different element layouts (row-major, Morton order and
Hilbert order). All multiplications were performed on square matrices of $2^{n}$ double precision
elements, for $n = 10, 11, 12$. The range of sizes is selected to expose the transition from
problems that fit entirely into the last-level cache on a single socket, through to memory-bound
application behavior. Clock frequency is varied either by fixing it to predetermined values of
1200MHz, 1800MHz or 2600MHz, or leaving it to vary in proportion to computational load, at the
discretion of the Linux \emph{ondemand} power governing mechanism. Multithreaded experiments were
controlled with the thread/core affinity settings of the GNU OpenMP library. Different
configurations are either denoted by \emph{s} where all threads are bound to the cores on a single
socket, or by \emph{d} where they are distributed evenly between two sockets. We also conducted
experiments with enabled hyperthreading, however, their outcomes proved to be only marginally
different from the configurations with the highest thread counts, and we therefore omitted them from
the paper.

\begin{table}
\caption{Experiment Configurations}
\label{tab:experiments}
\begin{center}
	\begin{tabular}{|c|c|c|c|}\hline
	Multiplication					&	Size	& Frequency	& Thread count\\\hline
	Row-major								& 10		& 1200 MHz	& 1s\\
	Morton order						& 11		& 1800 MHz	& 4s\\
	Hilbert order						& 12		& 2600 MHz	& 8s\\
													&				& ondemand	& 2d\\
													&				&						& 8d\\
													&				&						& 16d\\\hline
	\end{tabular}
\end{center}
\end{table}

\section{Results and Discussion}
\label{results}

Our exhaustive search of the parameter space described in Section \ref{methodology}
results in a set of $216$ sample points, each featuring a large number of
recorded time and energy measurements. In the interest of brevity, we therefore
begin this section by noting that generally, all results show execution time
varying in proportion to problem size, and inversely with clock frequency and thread
count.

Adding the row-major indexing cost of 1 multiplication and addition to the discussion
in Section \ref{sfcurves}, we can sort the costs of computing \textit{row-major (RM)},
\textit{Morton order (MO)}, and \textit{Hilbert order (HO)} indexing in ascending order with respect
to operation counts. Recorded execution times most notably reflect this by HO indexing giving the
consistently longest completion time, while the faster of the RM and MO schemes is determined by
sample configuration.

\subsection{Computational Performance}

\begin{figure}
\centering
\includegraphics[scale=0.65]{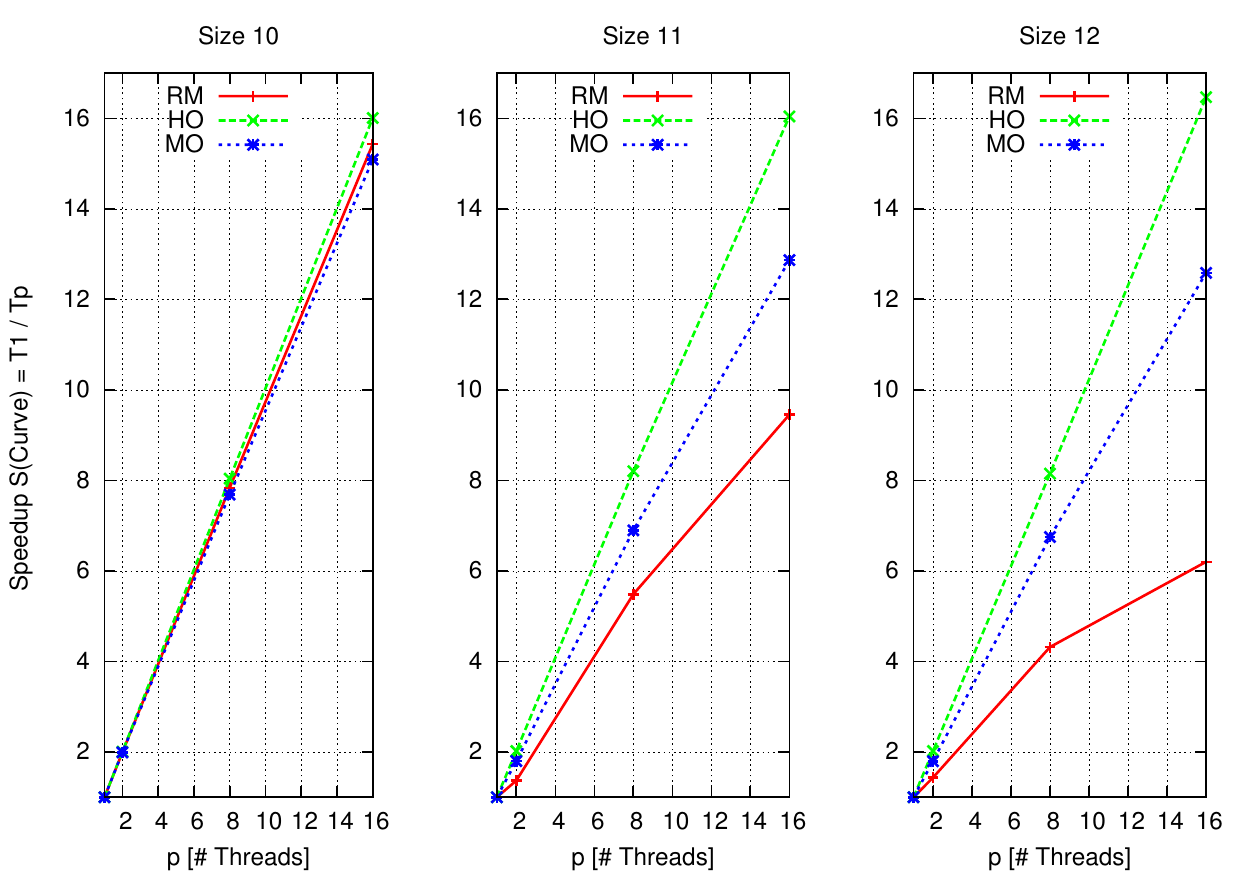}
\caption{Parallel speedup for all ordering schemes}
\label{fig:curves_speedup}
\end{figure}

\begin{figure}
\centering
\includegraphics[scale=0.65]{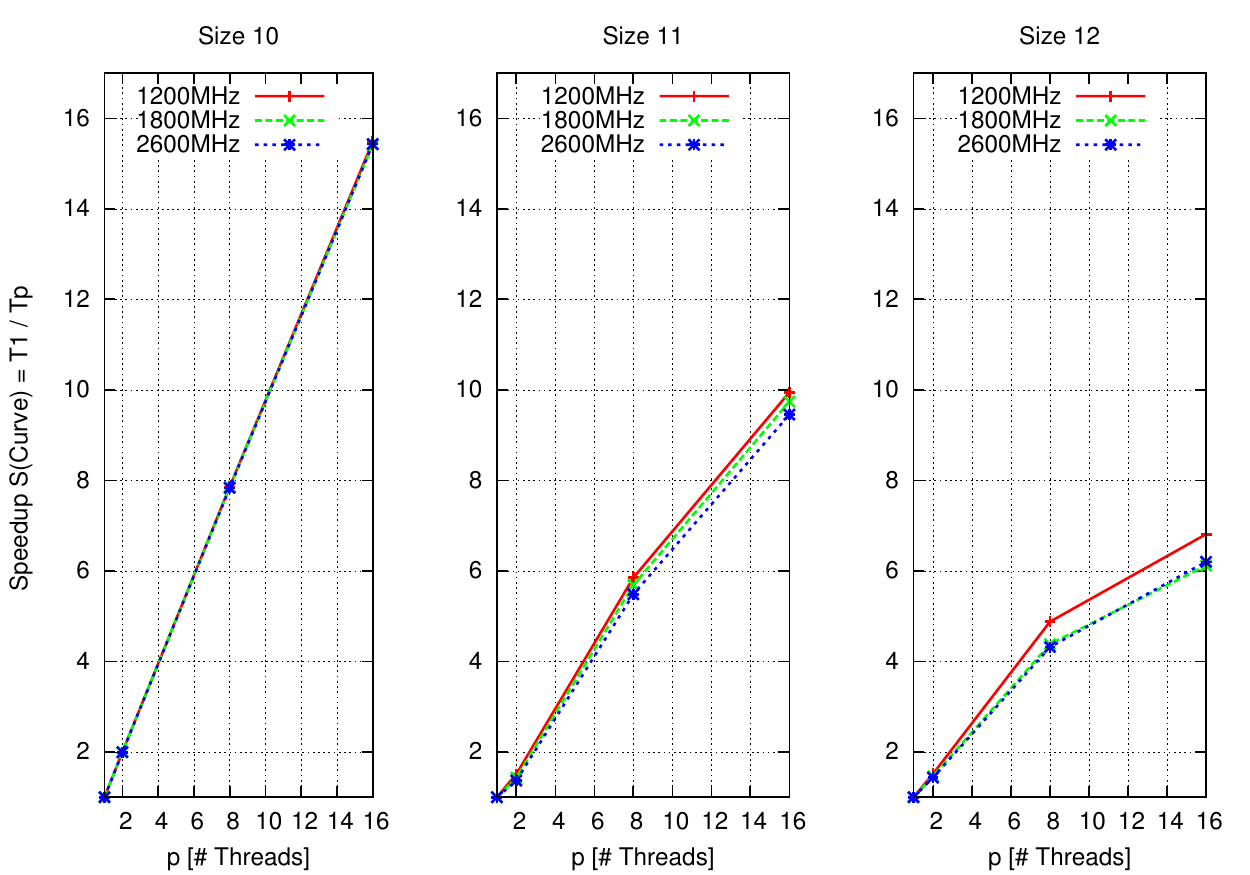}
\caption{Speedup of RM order with variable clock frequency}
\label{fig:rmj_speedup}
\end{figure}

Figure \ref{fig:curves_speedup} shows the parallel speedup of each ordering
scheme with variable core counts and problem sizes, and Figure \ref{fig:rmj_speedup}
shows the speedup of row-major ordering with variable clock frequencies and
problem sizes. Both figures show dual socket configurations, but similar
tendencies were evident in the single socket results, albeit less pronounced.

From the in-cache problem size \textit{$2^{10}$ (10)}, differences in parallel scalability are
barely distinguishable, demonstrating that the ordering of the memory
access pattern is insignificant, making the superior execution time of
the RM pattern favorable.

\begin{table}
\centering
\caption{Absolute execution times [s]}
\label{tab:absolute_times}
\begin{tabular}{|c|c|c|c|c|c|c|c|}
\hline
\multicolumn{2}{|c|}{{\bf RM}} & \multicolumn{3}{|c|}{Single Socket} & \multicolumn{3}{|c|}{Dual Socket}\\\hline
\multirow{2}{*}{Size} & F. in & \multirow{2}{*}{{\bf 1}} & \multirow{2}{*}{{\bf 4}} & \multirow{2}{*}{{\bf 8}} & \multirow{2}{*}{{\bf 2}} & \multirow{2}{*}{{\bf 8}} & \multirow{2}{*}{{\bf 16}} \\
& GHz & & & & & & \\
\hline
\multirow{4}{*}{10} & 1.2	& 7.2 & 1.8 & 0.9 & 3.6 & 0.9 & 0.4 \\
										& 1.8	& 4.8 & 1.2 & 0.6 & 2.4 & 0.6 & 0.3 \\
										& 2.6	& 3.3 & 0.8 & 0.4 & 1.6 & 0.4 & 0.2 \\
										& od	& 2.8 & 0.6 & 0.3 & 1.3 & 0.3 & 0.1 \\
\hline
\multirow{4}{*}{11} & 1.2	& 22.9 & 35.3 & 19.0 & 89.8 & 23.2 & 13.6\\
										& 1.8	& 108.4 & 28.0 & 14.7 & 75.8 & 19.0 & 11.1\\
										& 2.6	& 91.9 & 24.6 & 13.1 & 67.2 & 16.7 & 9.7 \\
										& od	& 84.5 & 22.7 & 12.6 & 63.6 & 14.9 & 9.6 \\
\hline
\multirow{4}{*}{12}	& 1.2	& 214.9 & 336.9 & 178.0 & 793.8 & 246.1 & 176.3\\
										& 1.8	& 996.2 & 276.9 & 164.1 & 680.3 & 227.2 & 162.8\\
										& 2.6	& 910.1 & 254.1 & 153.0 & 635.0 & 210.8 & 146.7\\
										& od	& 873.7 & 244.4 & 147.2 & 600.0 & 193.5 & 131.7\\
\hline
\hline
\multicolumn{2}{|c|}{{\bf MO}} & \multicolumn{3}{|c|}{Single Socket} & \multicolumn{3}{|c|}{Dual Socket}\\\hline
\multirow{2}{*}{Size} & F. in & \multirow{2}{*}{{\bf 1}} & \multirow{2}{*}{{\bf 4}} & \multirow{2}{*}{{\bf 8}} & \multirow{2}{*}{{\bf 2}} & \multirow{2}{*}{{\bf 8}} & \multirow{2}{*}{{\bf 16}} \\
& GHz & & & & & & \\
\hline
\multirow{4}{*}{10} & 1.2	& 13.6 & 3.4 & 1.7 & 6.8 & 1.7 & 0.8 \\
										& 1.8	& 9.1 & 2.3 & 1.1 & 4.5 & 1.1 & 0.5 \\
										& 2.6	& 6.2 & 1.6 & 0.8 & 3.1 & 0.8 & 0.4 \\
										& od	& 5.1 & 1.3 & 0.7 & 2.5 & 0.6 & 0.3 \\
\hline
\multirow{4}{*}{11} & 1.2	& 125.3 & 32.1 & 16.4 & 69.9 & 17.6 & 9.1 \\
										& 1.8	& 87.4 & 22.4 & 11.6 & 50.1 & 12.5 & 6.6 \\
										& 2.6	& 63.5 & 16.4 & 8.5 & 35.3 & 9.2 & 4.9 \\
										& od	& 51.6 & 13.8 & 7.4 & 28.7 & 8 & 4.5 \\
\hline
\multirow{4}{*}{12} & 1.2	& 1011.3 & 259.1 & 132.3 & 571.5 & 144.4 & 74.6 \\
										& 1.8	& 706.2 & 180.8 & 93.2 & 410.6 & 103.8 & 54.1 \\
										& 2.6	& 514.6 & 132.9 & 68.3 & 286.8 & 76.2 & 40.8 \\
										& od	& 419.7 & 113.4 & 61.1 & 236.8 & 65.9 & 37.0 \\
\hline
\hline
\multicolumn{2}{|c|}{{\bf HO}} & \multicolumn{3}{|c|}{Single Socket} & \multicolumn{3}{|c|}{Dual Socket}\\\hline
\multirow{2}{*}{Size} & F. in & \multirow{2}{*}{{\bf 1}} & \multirow{2}{*}{{\bf 4}} & \multirow{2}{*}{{\bf 8}} & \multirow{2}{*}{{\bf 2}} & \multirow{2}{*}{{\bf 8}} & \multirow{2}{*}{{\bf 16}} \\
& GHz & & & & & & \\
\hline
\multirow{4}{*}{10} & 1.2	& 90.0 & 22.3 & 11.2 & 45.0 & 11.2 & 5.6\\
										& 1.8	& 59.5 & 14.8 & 7.4 & 29.8 & 7.4 & 3.7 \\
										& 2.6	& 41.4 & 10.2 & 5.1 & 20.6 & 5.1 & 2.5 \\
										& od	& 32.6 & 8.4 & 4.4 & 16.3 & 4.2 & 2.2 \\
\hline
\multirow{4}{*}{11} & 1.2	& 865.4 & 212.7 & 105.8 & 440.8 & 108.6 & 53.8\\
										& 1.8	& 583.3 & 143.1 & 71.2 & 298.2 & 73.1 & 36.2 \\
										& 2.6	& 409.9 & 100.2 & 50.0 & 203.0 & 49.9 & 25.5 \\
										& od	& 324.4 & 83.6 & 43.0 & 160.7 & 41.4 & 21.6 \\
\hline
\multirow{4}{*}{12} & 1.2	& 7661.9 & 1887.1 & 939.9 & 3890.0 & 962.3 & 477.5\\
										& 1.8	& 5155.4 & 1267.4 & 632.2 & 2629.0 & 649.4 & 321.9\\
										& 2.6	& 3619.0 & 886.6 & 441.3 & 1792.5 & 444.1 & 219.8 \\
										& od	& 2861.4 & 734.5 & 382.4 & 1410.3 & 366.6 & 193.1 \\
\hline
\end{tabular}
\end{table}

Problem sizes 11 and 12 clearly show the transition to memory-bound behavior
for the RM and MO schemes: 3 double-precision square matrices of dimension \textit{$2^{11}$ (11)}
amount to total problem sizes of 96\textit{MB} and 348\textit{MB} respectively, and from the
straightforward $n^3$ matrix multiplication routine, we can infer that access to the A and B
matrices dominate memory access cost. The aggregate amount of last-level cache in our
dual-socket configuration is 40\textit{MB}, and the degradation is expected. As
the speedup curves are relative to a single-thread baseline per curve, the
effect on absolute speed is not visible in the figure, but as Table \ref{tab:absolute_times}
shows, the favorable access pattern of the MO overtakes RM
in terms of absolute performance.

As the extreme case of trading locality for computation, the scalability and
execution time of HO displays the interesting property that while its absolute
execution time is an order of magnitude higher than RM and MO, its speedup
curve shows slight superlinearity with increasing parallelism, specifically,
at 8 threads for problem size 11, and both 8 and 16 threads for size 12.

We mainly attribute the favorable scalability of the HO experiments to the 
fact that the additional computation it requires parallelizes trivially,
which suggests that the HO computations would become memory bound
in a similar fashion to the MO computations, if the test platform supported
a sufficient number of threads. This does not account for superlinear speedup,
however, so a further examination aimed to establish whether this is a
consequence of deviations in the measurements of the single-threaded baseline,
or a consequence of the data locality properties of the Hilbert curve.

As the results show that our test platform has an insufficient number of cores
to precipitate memory bound behavior from the HO experiments, we instead
investigate cache utilization using the \emph{cachegrind} module of the
\emph{Valgrind} instrumentation framework. It allows matching of
memory hierarchy effects to specific locations in the source program.
An instrumentation overhead of an approximate factor $100$ made it prohibitive
to fully run our largest scale computation to completion in this manner,
but an estimate can be obtained by restricting the ZO and HO codes to complete
a small number of rows in the output matrix, thereby ensuring that several complete
traversals of one entire input matrix have been performed, to elicit effects
of any significantly different locality properties of its encoding.

Performing this additional experiment for 5 rows near the middle of the C
matrix in a size 12 problem resulted in a total of $16.78 \times 10^6$
last-level data read misses for HO compared to $17.06 \times 10^6$ for MO,
suggesting that while the superior data locality is far from significant
enough to amortize the additional computation cost of Hilbert indices on
our test system, it is a measurable effect that makes its applications more
amenable to parallelization. 

\subsection{Energy Consumption}

\begin{figure*}
\centering
\includegraphics[scale=0.75]{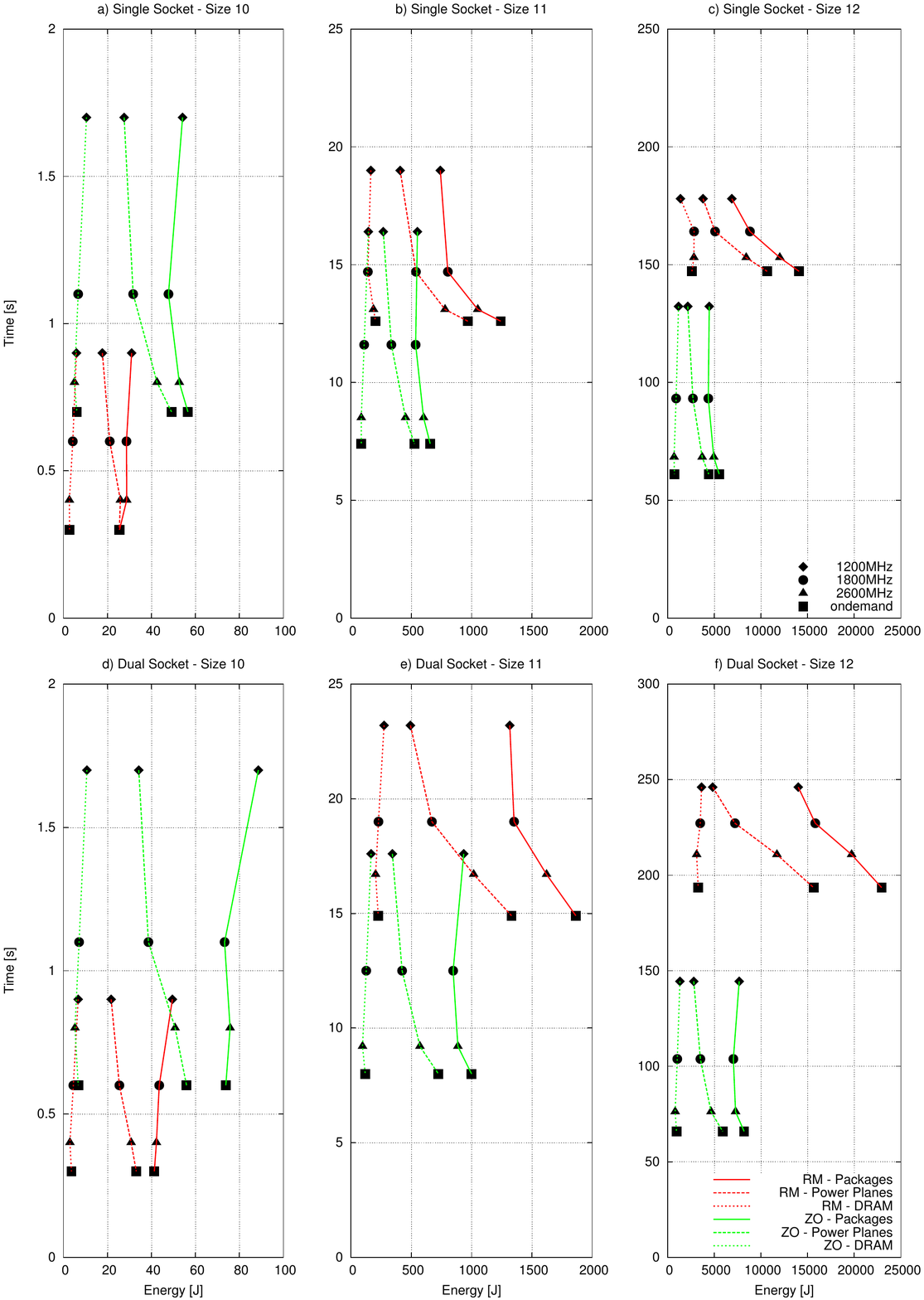}
\caption{Energy and time samples from experiment configurations 8s and 8d}
\label{fig:energy_vs_time}
\end{figure*}

The RAPL register results collected from each run are displayed in
Figure \ref{fig:energy_vs_time}, which plots total energy \emph{vs.}
execution time. Individual points represent particular energy figures, with dashed
and solid lines grouping points by the processor feature they represent, and
the 4 points of each line representing tests at different clock frequencies.

\emph{Package} curves represent the sum of energy from the individual
packages situated on each socket. \emph{Power plane} curves reflect the portion
of the package total that is due to the processing cores. \emph{DRAM} curves are
the sum of energy consumption from the memory modules attached to both sockets;
these are reported in sum for all configurations, because the virtual memory
abstraction of the operating system prevents programs from explicitly controlling
the placement of the memory allocations that belong to a given application.

The samples from the 8d and 8s configurations are representative of tendencies
throughout our results, but were selected because these configurations
permit comparisons of identical thread counts distributed on 1 and 2 sockets.
Results are shown for the RM and MO patterns only, as the computational overheads
of the HO cases are substantially larger, and would therefore require a figure scale
which would complicate comparisons.

Subfigures \emph{c)} and \emph{f)} show the largest problem size, where the
computation is memory bound, making the locality improvement of the MO most significant.
The RM results are interesting in these cases, because they show that while speed
advantages are attained at higher clock frequencies, the impact in energy consumption
is greater; a similar tendency can also be seen in subfigures \emph{b)} and \emph{e)}.
Noting that the effect appears from frequencies 1800MHz and higher, it appears as a
consequence of raising the processing rate higher than the memory bus clock, which
operates at 1600MHz. Thus, the common assumption that optimal execution
speed can be equated with optimal energy efficiency must be refined in the case of
memory-bound computations, as the attainable speed improvements in this region 
come at an energy cost that is disproportionate to their gains. Note also that the
MO curve does not equally saturate the memory system, and continues to attain
improvements with rising frequency.

Subfigures \emph{a)} and \emph{c)} display the in-cache problem size 10, where
the order of data element access is practically insignificant, making the RM indexing the faster
implementation, and this translates
to lower energy consumption. We can also note that the in-cache problem size displays
another effect: the RM figures for powerplane and package grow closer with rising clock
frequencies, suggesting that the powerplane takes up an increasing fraction of the package
power budget with increasing computational speed. This can be contrasted with the memory
bound computations, where the package energy consumption follows that of the powerplane,
suggesting increasing loads on both the processing cores and their shared on-chip resources.

DRAM energy consumption is nearly constant, except for a slight shift towards lower total
energy usage for higher frequencies where computations complete faster. The most significant
observation from these measurements is that total DRAM power consumption is small in comparison 
to that of the processing cores' consumption, differing by factors close to 4 for high frequencies.

Contrasting single socket and dual socket results, we observe that the difference in execution
time becomes quite noticeable for large problem sizes, resulting in a difference of nearly 50
seconds between subfigures \emph{c)} and \emph{f)}. This reflects that the synchronization overhead
of the implicit barrier following a straightforward parallel OpenMP loop implies a greater
latency when threads are scattered across multiple sockets, and the energy impact is most notable
at low frequency.

The exact consequences of enabling the ondemand power governor mechanism are difficult
to establish without more detailed traces of run time behavior, but already from the presented
results, we note that our test system processors feature Intel's Turbo Boost performance
enhancements, which places the system at liberty to selectively overclock individual cores when
doing so does not exceed the package thermal design point. This allows it to produce superior
run times compared to results from maximal fixed frequency settings, but it consistently
makes energy efficiency deteriorate for out-of-cache problem sizes.

We took samples from our power estimates and compared them to the full system power measurements
obtained from the Yokogawa WT210. In our test system, the memory and the two CPUs account for
approximately $38\%$ of the total system consumption when all cores are utilized.

For completeness, we also performed comparisons with ATLAS. As expected, the ATLAS library
outperformed our multiplications by an order of magnitude, but at the cost of a one-time investment
of a two hour auto-tuning process.

\section{Related Work}
\label{relwork}
Space-filling curves have been studied extensively throughout the years, and have found use in several
application areas such as heuristics for combinatorial problems, data-bases and geographical
information systems as well as signal processing. Bader \emph{et al.} \cite{BaderMM} present an
algorithm for cache oblivious dense matrix-matrix multiplication based on the Peano curve
\cite{Peano90}. It uses a recursive block multiplication scheme to exploit the recursive nature of
the Peano curve, and is asymptotically optimal in the number of cache misses for an ideal cache.
Efficient parallel implementations have been developed \cite{Bader08, Heinecke08}, and an extension
for sparse matrix-matrix multiplications exists \cite{Bader09}. In computer graphics, space-filling
curves are used as scanning orders for images, preserving some of the spatial information of the
scanned multi-dimensional space within the single dimension of the scan. Witten \emph{et al.}
\cite{Witten} used this technique to construct a dither with a low cumulative error over large and
small regions of the image, whereas Ansari \emph{et al.} \cite{Ansari} and Yang \emph{et al.}
\cite{Yang} employ the Peano scan \cite{Stevens} for clustering highly correlated data before
compression, and therefore achieving a higher compression ratio. A thorough treatment of space-filling
curves along with their applications can be found in Michael Bader's book \cite{bader13}.

Over the last years there has been considerable research into reducing energy consumption of
workloads through the help of hardware counters. Ge \emph{et al.} \cite{Ge} use them to periodically
collect events such as retired instructions, L1/L2 data cache accesses and memory data accesses for
identifying on-/off-chip phases in workloads and adjusting the CPU's frequency with DVFS such
that energy usage is minimized while performance retained. Hsu \emph{et al.} \cite{Hsu} make similar
predictions based on MIPS (millions of instructions per second). Huang \emph{et al.} \cite{Huang}
use the CPU's decoder/dispatch stall cycles in addition to L2 cache misses, and stall cycles
due to branch misprediction. Porterfield \emph{et al.} \cite{Porterfield} use the RAPL interface to
dynamically throttle the number of threads to reduce the energy consumption in OpenMP programs.

\section{Conclusions and Future Work}
\label{conclusion}

In this paper, we studied the energy and locality effects of the Morton and Hilbert space-filling
curves, with application to dense matrix-matrix multiplication. We showed
that energy consumption is only exclusively proportional to the execution time when the computation
is CPU bound. When memory traffic begins to dominate computation, increasing the frequency results
in higher energy consumption in exchange for diminishing run time improvements. The results 
also revealed that the energy spent accessing DRAM is small and constant compared to the amount
spent by caches and CPU. We conclude that the gap between memory and CPU performance has significant
impact on energy efficiency, making improvements in the utilization of the memory hierarchy
as important as minimizing execution time.

Selecting appropriate data structures is one way to improve memory hierarchy utilization, and we
investigated the application of Morton and Hilbert orders to matrix multiplication.
We found that while both orders improved spatial locality, Morton order results in
practical improvements over naive approaches on our test platform, while the greater computational
requirements of the Hilbert ordering render it impractical.

Although the locality properties of the Hilbert ordering were eclipsed by the associated indexing
cost, we observed that its locality effect offers a moderate improvement over that of the Morton
order. The additional computational cost of Hilbert ordered indexing amounts to simple bitwise
register manipulations. An interesting direction for future work would be to investigate the
benefit of dedicated hardware support for the required operations, as this would greatly reduce
the overhead.

\section*{Acknowledgement}
The authors would like to thank Juan Manuel Cebri\'{a}n Gonz\'{a}lez for his kind assistance
with the instrumentation equipment, and gratefully acknowledge the support of the NOTUR project.

\end{document}